\documentclass{aa}  

\usepackage{graphicx}
\usepackage{txfonts}
\usepackage{wasysym}
\usepackage{subcaption}
\usepackage{graphicx}
\usepackage{hyperref}

\begin{document}

   \title{Turbulence destroys thermal lobes around Mars-sized planetary embryos}
   \titlerunning{Turbulence destroys the thermal lobes}

   \author{R. O. Chametla 
          \inst{1}
          \and 
          A. Moranchel-Basurto\inst{1}
          \and 
          F. J. Sánchez-Salcedo\inst{2}
          }
   \authorrunning{R.~O.~Chametla et al.}
   \institute{%
            Charles University, Faculty of Mathematics and Physics, Astronomical Institute,
            V Hole\v{s}ovi\v{c}k\'ach 747/2, 180 00,
            Prague 8, Czech Republic\\
            \email{raul@sirrah.troja.mff.cuni.cz}
             \label{UKarlova}
         \and
            Instituto de Astronomía, Universidad Nacional Autónoma de México, 
            Apt.~Postal 70-264, C.P.~04510, Mexico City, Mexico
             \label{UNAM}
             }

   \date{Received XXX; accepted YYY}
   
  \abstract
   {The release of heat by a planetary embryo modifies the local density perturbations, forming thermal lobes in its vicinity, and thereby altering the torque exerted by the disk on the embryo. In laminar disks, these thermal torques can dominate the disk-embryo interaction, rendering the classical Lindblad and corotation torques largely subdominant.
   }
   {The aim of this work is to investigate how turbulence driven by the magnetorotational instability affects the thermal lobes formed around a planetary embryo, and to analyze the resulting torque acting on the embryo.
   }
   {We evaluate the thermal torques exerted on a planetary embryo of mass $M_p=0.33M_{\mars}$ (with $M_{\mars}$ is the mass of Mars) and on a planetary core with mass $M_{p}=1M_{\oplus}$, each embedded in a turbulent gaseous protoplanetary disk, by means of high-resolution three-dimensional magnetohydrodynamics simulations that include thermal diffusion and an initially toroidal magnetic field. The magnetic field strength is characterized by the plasma $\beta$ plasma parameter with $\beta\in\{50,1000\}$. We consider two values for the luminosity of the planetary embryo: $L=0$ (cold thermal lobes) and $L=L_c$ (hot thermal lobes), where $L_{c}$ represents
   the critical luminosity. For the $0.33M_{\mars}$ embryo, $L_c=7.8\times10^{25}\mathrm{ergs}\,\mathrm{s}^{-1}$ when it orbits a Sun-like star at a distance $r_p=5.2\,\mathrm{au}$.}
   {We find that, even in the presence of a weak magnetic field and irrespective of the luminosity, for both planetary masses, the development of turbulence in the disk (which takes between 1.5 to 3 orbital periods) completely disrupts the thermal lobes. 
   As a result, the torque acting on both the planetary embryo and the Earth-mass core displays a strongly oscillatory behavior.
   This suggests that planets with masses in the range $0.03M_{\oplus}\lesssim M_{p}\lesssim 1M_{\oplus}$ 
   experience stochastic migration, as expected in turbulent disks. }
   {Thermal torques become inefficient in turbulent regions of protoplanetary disks, such as outside the dead zone, in both the inner and outer disk regions where the magnetorotational instability operates.}


   \maketitle
%

\section{Introduction}

In recent years, thermal torques on low-mass planets and planetary embryos embedded in regions of the gaseous protoplanetary disks where the flux is laminar \citep{Lega_etal2014,Masset2017,BLl_etal2015,ChM2021} have emerged as a possible solution to the problem of fast inward migration \citep[see][for a review]{Paar2023}. Thermal torques on a planet are due to the asymmetric gas density structures formed in the vicinity of the planet, which result from thermal diffusion in the disk. When a planet accretes pebbles, it releases heat into the environment, producing the formation of low-density, asymmetric hot lobes in the disk. These lobes, in turn, produce a torque on the planet \citep[the heating torque;][]{Masset2017}. 
If the planet does not accrete pebbles, thermal diffusion produces the formation of high-density cold lobes in the coorbital region of the planet \citep[the cold thermal torque;][]{Lega_etal2014,Masset2017}. 
Both heating and cold thermal torques can exceed the Lindblad and corotation torques by up to an order of magnitude, such that the Lindblad and corotation torques become negligible. Heating thermal torques can slow down, or even reverse, the inward migration of growing low-mass planets \citep{BLl_etal2015}. 

Using global disc models, \citet{Guilera2019} showed that thermal
torques can modify both migration maps and 
planet formation tracks, i.e.  
the evolution of the planetary mass as a function of semi-major axis
\citep[but see also][]{Baumann2020}. In a more comprehensive framework
of mass growth by pebble accretion, \citet{Guilera2021} found that thermal torques play a significant role in shaping planet formation tracks 
in low-viscosity disks, particularly when the disks are both massive and metal-rich. Under these conditions,
planets forming beyond the ice line may undergo substantial outward migration.

Thermal forces can also excite eccentricity and inclination.
In a model that additionally follows the evolution of eccentricity and inclination for
a population of pebble-accreting embryos, \citet{cornejo2023} 
reported that the inclusion of thermal torques leads to a markedly different dependence of the final embryo mass spectrum on disk metallicity. 
The excitation of eccentricity by thermal torques may also have important implications for the trapping and formation of planets in pressure bumps \citep{Chrenko2023,Pierens2024,VRMM2024}.

All studies of thermal torques to date have been conducted under the assumption of an unmagnetized, laminar disk.
However, there exist regions of protoplanetary disks in which turbulence is expected to dominate the transport of angular momentum. In such turbulent environments, it has been shown that the migration of planetary embryos can become stochastic \citep{NP2004,Nelson2005}. This occurs, for instance, outside the dead zone,
where strong magnetic coupling enables turbulence driven by the magnetorotational instability \citep[MRI;][]{BH1991}.
In addition, hydrodynamical instabilities may also lead to turbulence even within the dead zone \citep[see][]{LU2019}.
In this work, we assess whether thermal torques remain effective in driving the migration of low-mass planets or planetary embryos in turbulent regions of the disk.
To address this, we conduct high-resolution MHD simulations of a planetary embryo embedded in a thermally diffusive, stratified protoplanetary disk threaded by a toroidal magnetic field. 

The paper is laid out as follows. Section \ref{sec:equations} describes the MHD equations, the numerical setup, and the code used in this work. In Section \ref{sec:results} we present our results. A brief discussion of possible improvements in more complex models is presented in Section \ref{sec:complex_mod}.  
Finally, concluding remarks can be found in Section \ref{sec:conclusions}. 

\section{Physical and numerical setup}
\label{sec:equations}
We consider a 3D magnetized, inviscid gas disk whose dynamical evolution is described by the MHD equations:

\begin{figure}
\includegraphics[width=0.4\textwidth]{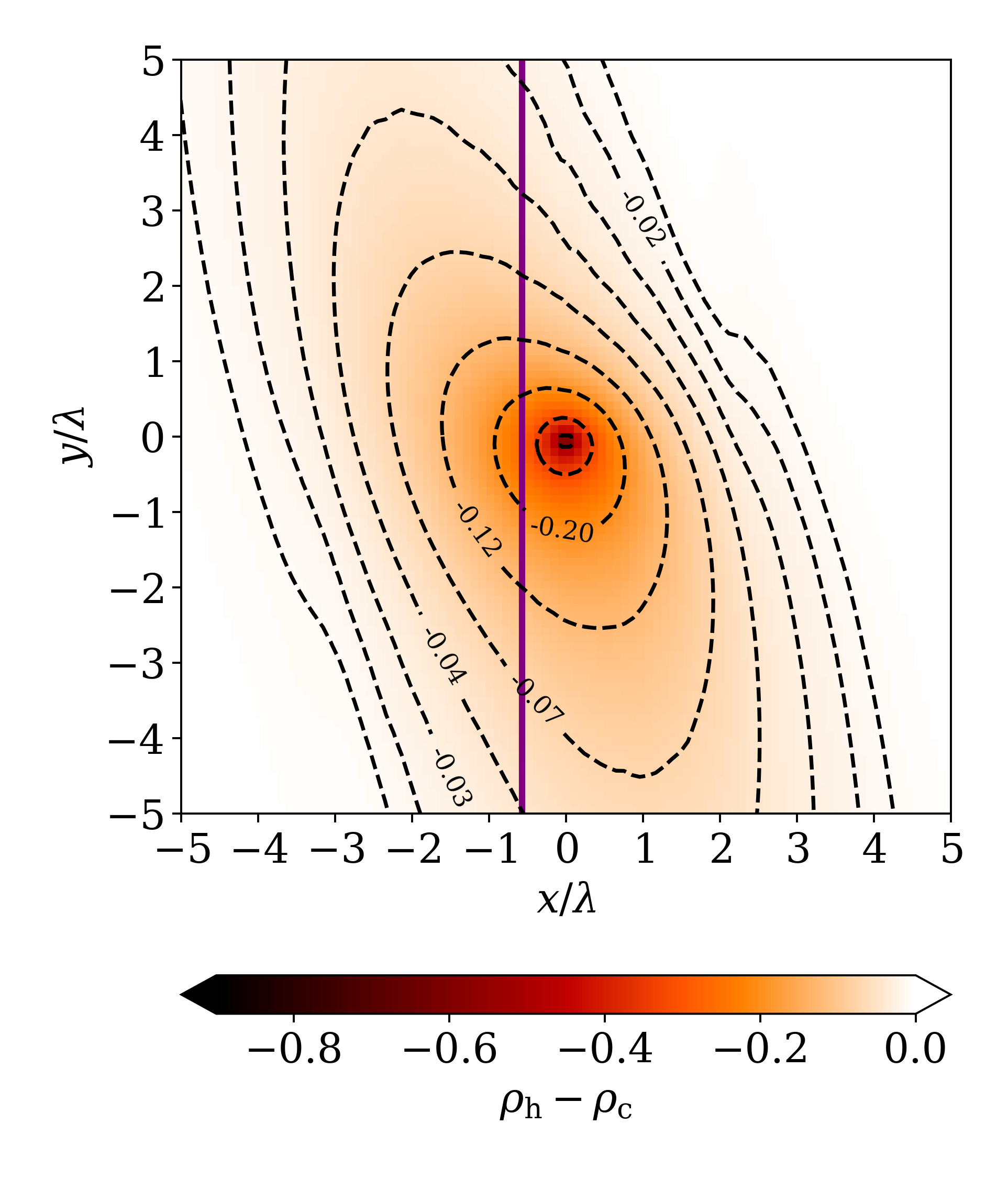}
 \caption{Perturbation of density arising from heat release obtained by subtracting a hot run ($L=L_c$) and a cold run ($L=0$) for the case $\beta=50$ at $t=1$ orbit. The perturbation is integrated over colatitude and normalized to $\gamma(\gamma-1)L/\chi c_s^2$. The purple solid line indicates the corotation radius.}
\label{fig:lobes}
\end{figure}

\begin{figure}
\includegraphics[width=0.4853\textwidth]{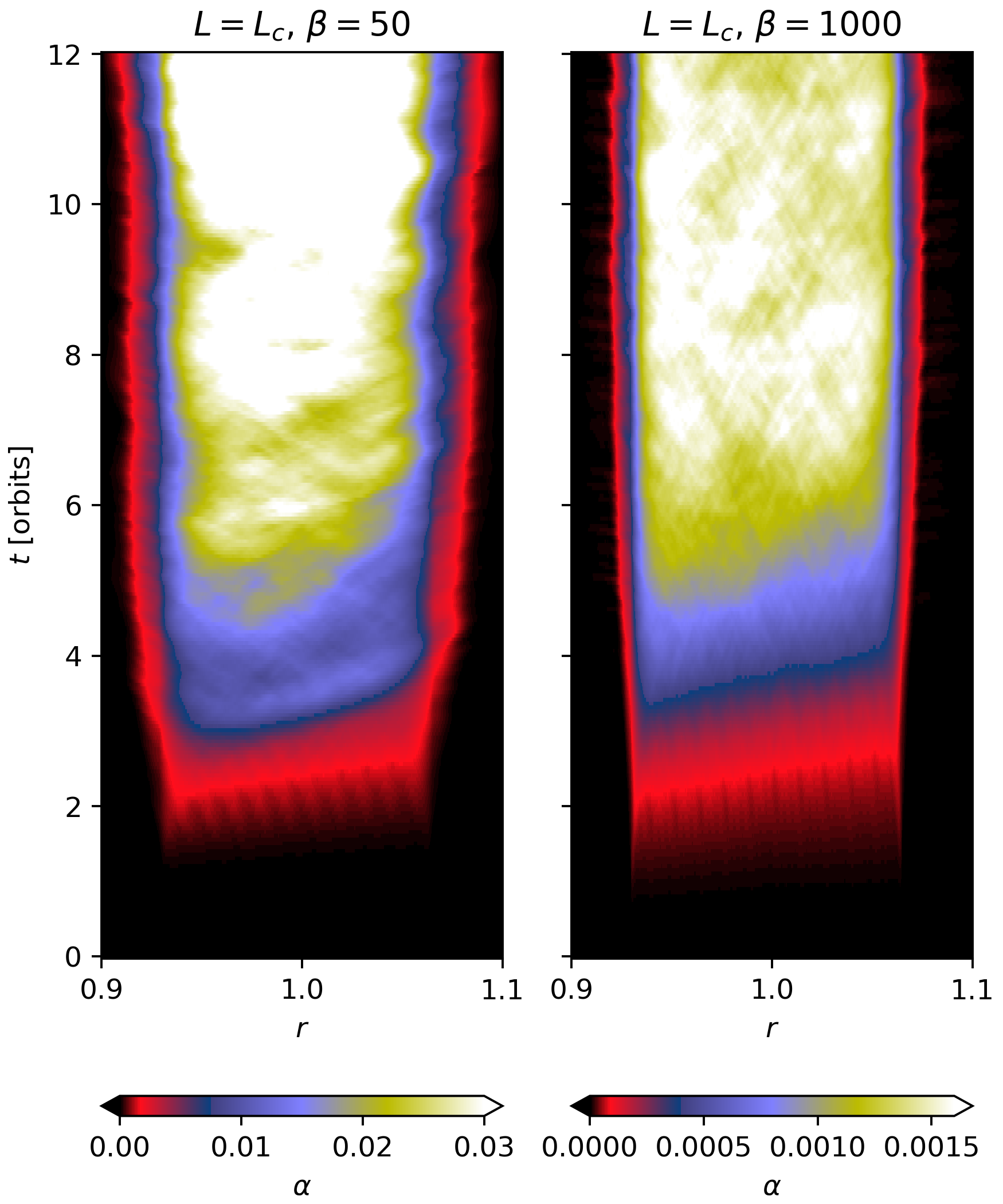}
 \caption{Temporal evolution of $\alpha$, defined by Eq. (\ref{eq:alpha}), for the runs with $L=L_c$ and $\beta=50$ 
 (left panel) and $\beta=1000$ (right panel).}
\label{fig:alpha}
\end{figure}

\begin{figure}
\includegraphics[width=0.4853\textwidth]{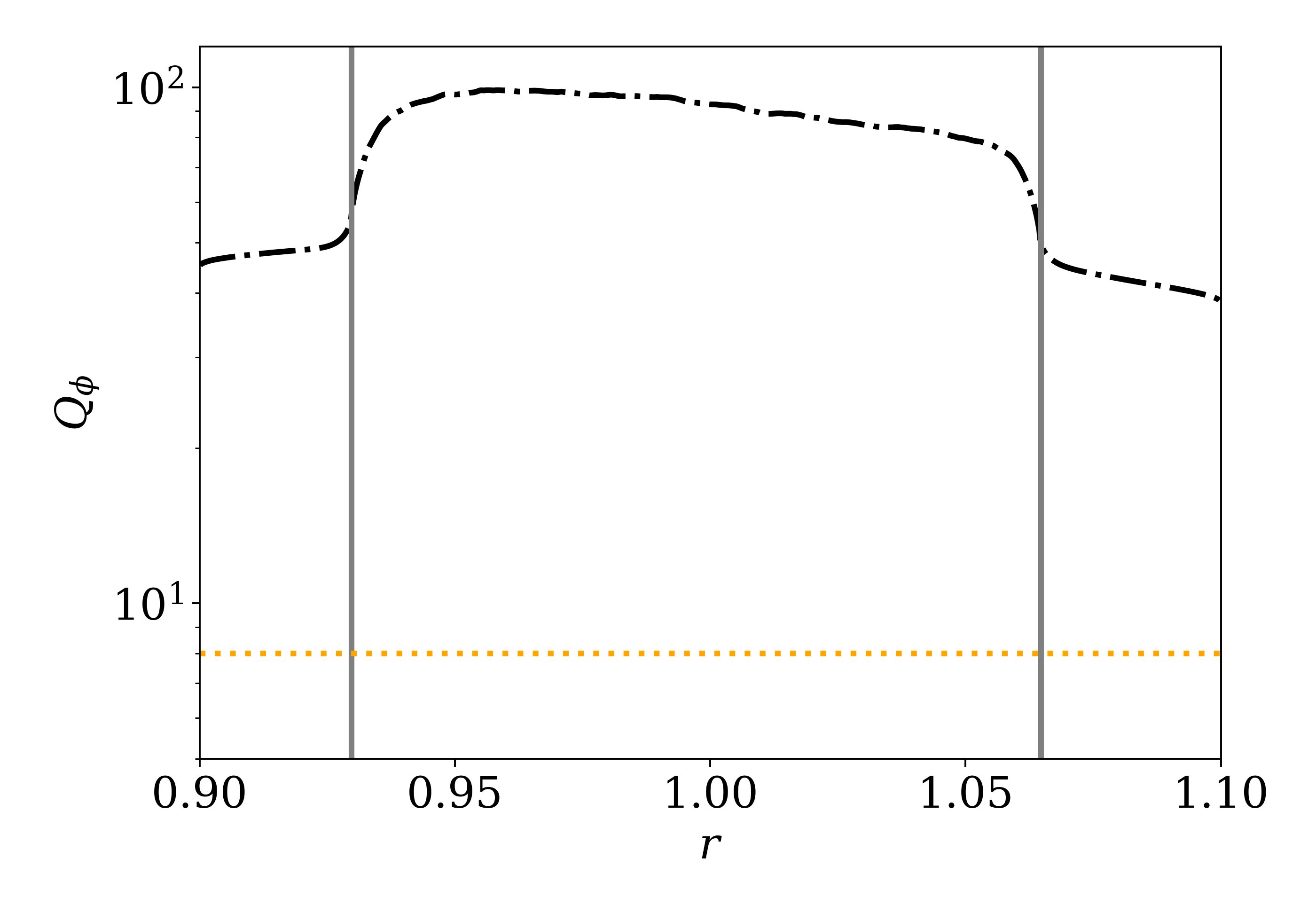}
 \caption{Quality factor averaged over space and time. The orange dotted line shows the eight-cell limit. The vertical gray lines delineate the buffer zones.}
\label{fig:qf}
\end{figure}

\begin{align}
\partial_t\rho+\nabla\cdot(\rho \mathbf{v})&=0,
 \label{eq:gas_cont}\\
\partial_t(\rho\mathbf{v})+\nabla\cdot(\rho\mathbf{v}\otimes\mathbf{v}+p\mathbf{I})&= -\rho\nabla\Phi+\mathbf{J}\times\mathbf{B},
 \label{eq:gas_mom}\\
\partial_te+\nabla\cdot(e\mathbf{v})&=-p\nabla\cdot\mathbf{v}-\nabla\cdot\mathbf{F}_\mathrm{H} + S_{p}
 \label{eq:gas_energy},\\
\frac{\partial \mathbf{B}}{\partial t}&=\nabla\times(\mathbf{v}\times\mathbf{B}),
    \label{eq:induc}
\end{align}
where $\rho$ and $\mathbf{v}$ denote the gas density and velocity, respectively, and $\mathbf{J}$ and $\mathbf{B}$ the current density and magnetic field. $\mathbf{I}$ denotes the unit tensor and $\Phi$ the gravitational
potential. For the gravitational potential of the planetary embryo we adopt a softening length comparable to our mesh resolution. 
For the pressure $p$, we consider the
equation of state
\begin{equation}
p=(\gamma-1)e,
 \label{eq:pressure}
\end{equation}
where $e$ is the internal energy density and $\gamma=7/5$ the gas adiabatic index.

In Eq. (\ref{eq:gas_energy}), $S_{p}$ is a source term arising from the energy release of the planetary embryo defined as
\begin{equation}
S_p=L\delta(\mathbf{r}-\mathbf{r}_p)
    \label{eq:source_term}
\end{equation}
with $L$ and $\mathbf{r}_p$ the luminosity and the location of the planet, respectively, and $\delta$ the Dirac’s delta function. While $\mathbf{F}_\mathrm{H}$ is the heat flux, given by
\begin{equation}
\mathbf{F}_\mathrm{H}=-\chi\rho\nabla\left(\dfrac{e}{\rho}\right),
    \label{eq:heatflux}
\end{equation}
where $\chi$ is the thermal diffusivity. 
This advection-diffusion approximation has been commonly adopted in previous studies
\citep[e.g.,][]{Masset2017,VRM2020,ChM2021,Chame2025}. Other authors instead employ the flux-limited diffusion scheme within a two-temperature framework \citep{BLl_etal2015}. In the regime considered here, however,
both approaches describe radiative transport in the diffusion limit, and therefore we do not expect significant differences in the resulting thermal torques.

\begin{figure*}
     \begin{subfigure}[b]{0.3\textwidth}
         \centering
         \includegraphics[width=\linewidth]{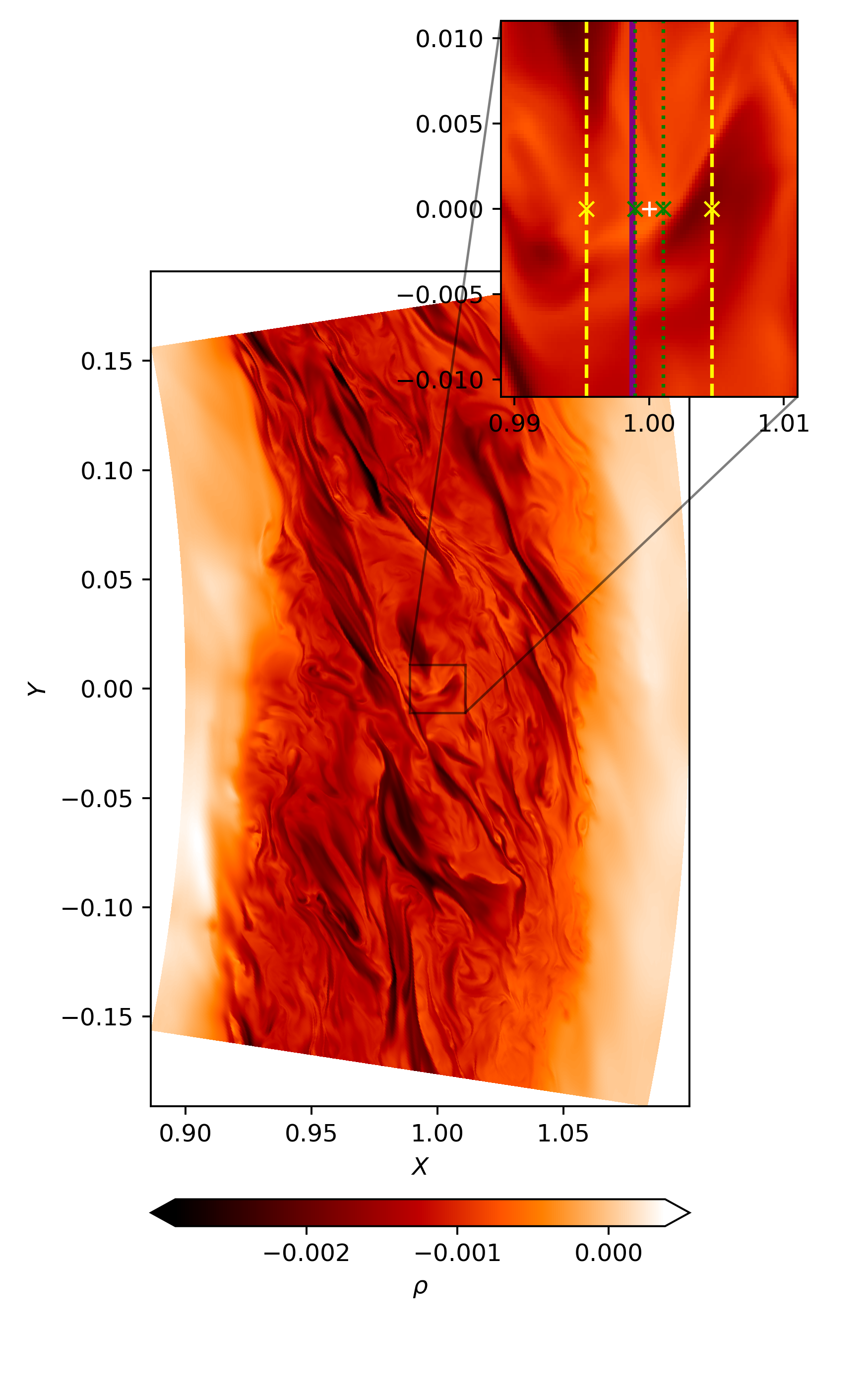}
         \caption{$L=0$, $\beta=50$}
         \label{fig: Logrestic Regression}
     \end{subfigure}
     \hfill
     \begin{subfigure}[b]{0.3\textwidth}
         \centering
         \includegraphics[width=\linewidth]{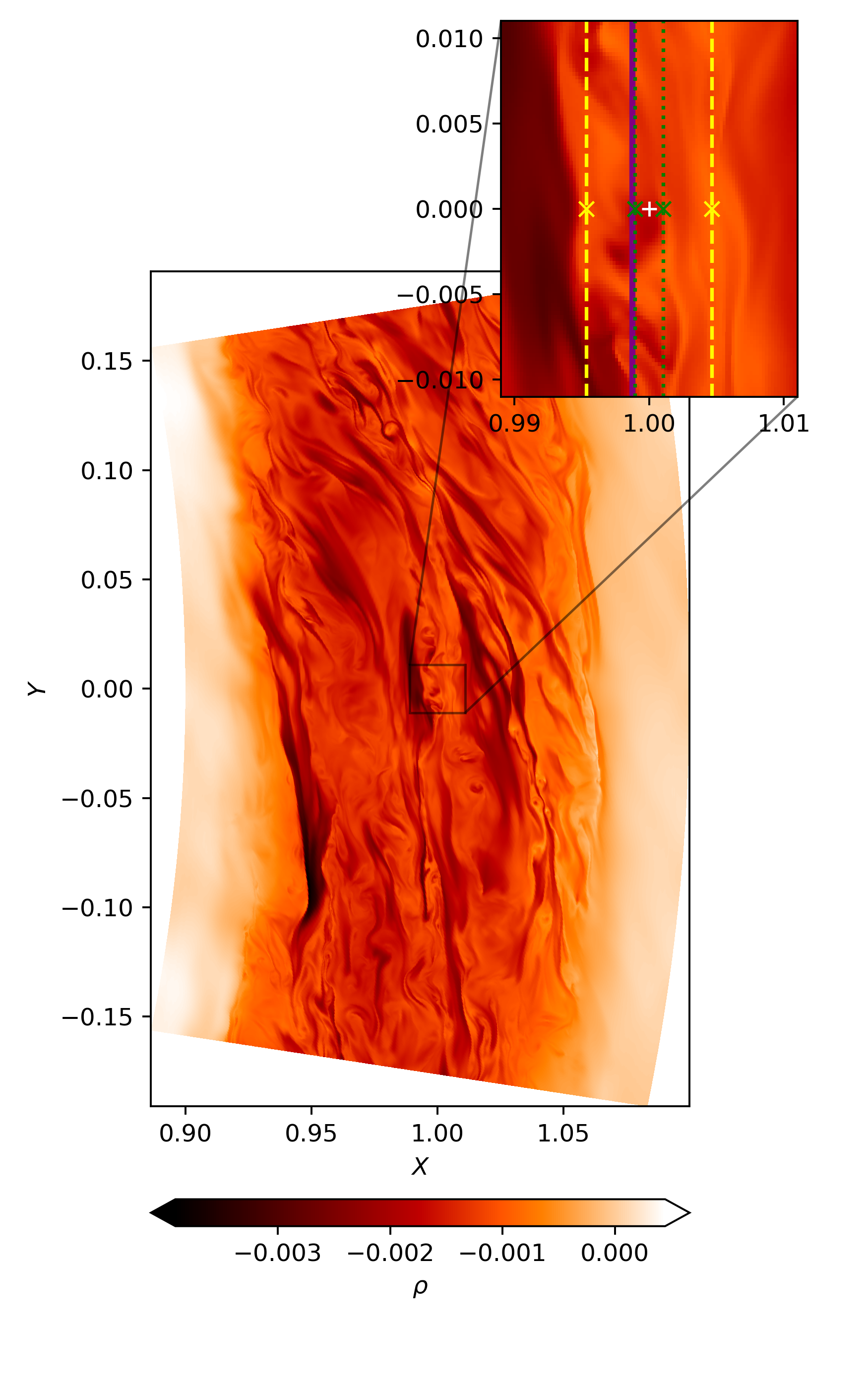}
         \caption{$L=L_c$, $\beta=50$}
         \label{fig: Naive Bayes}
     \end{subfigure}
     \hfill
     \begin{subfigure}[b]{0.3\textwidth}
         \centering
         \includegraphics[width=\linewidth]{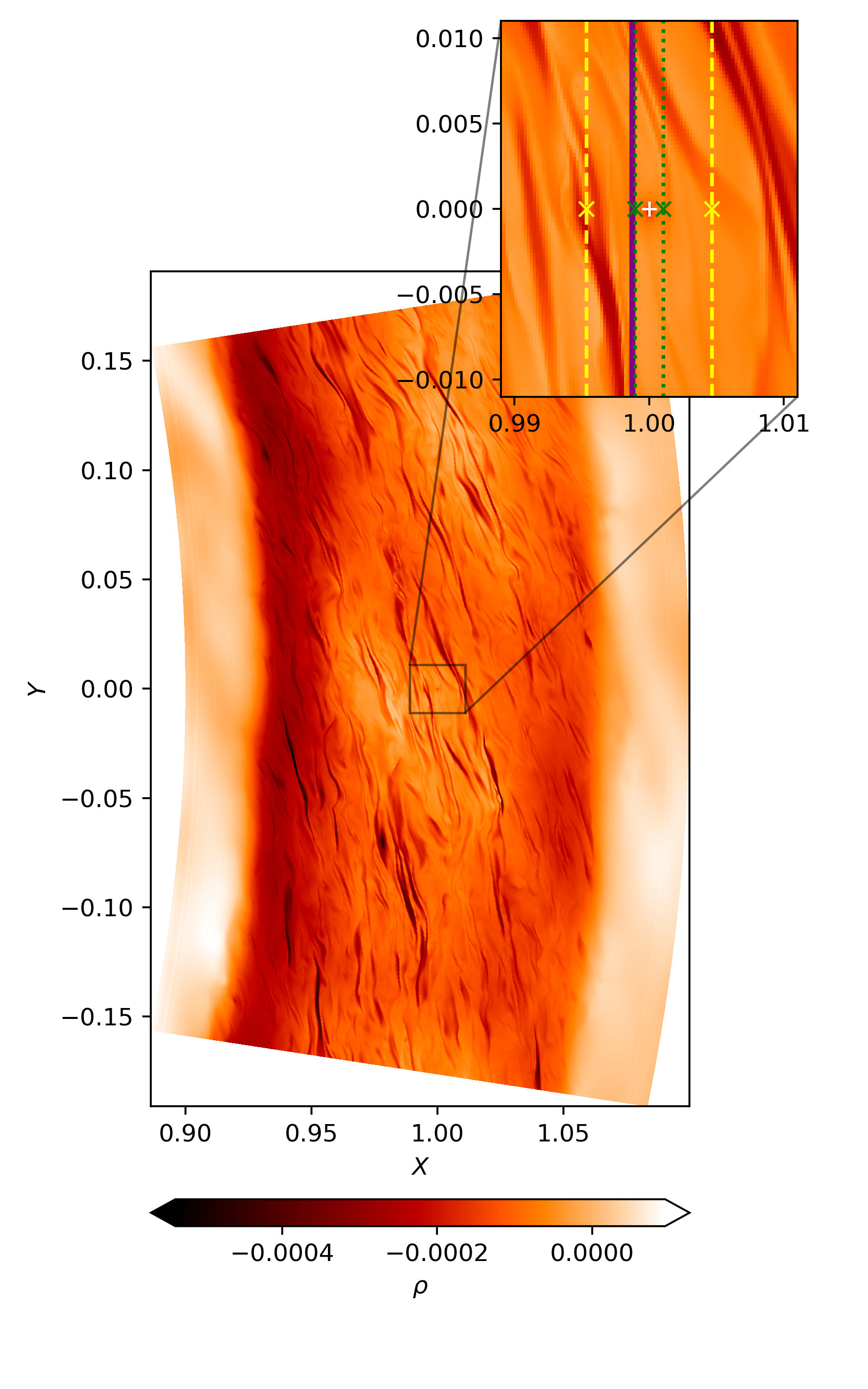}
         \caption{$L=L_c$, $\beta=1000$}
         \label{fig: SVM}
     \end{subfigure}
        \caption{Gas density (logarithmic scale) at the midplane at $t=11$ orbits. The zoomed regions correspond to an area similar to that shown in Fig. \ref{fig:lobes} around the planetary embryo. In these regions, no low-density structures resembling hot lobes are observed. The purple solid lines indicate the corotation radii.  The yellow dashed and green dotted lines mark the locations of the magnetic resonances for $\beta=50$ and $\beta=1000$, respectively. The white plus symbol denotes the position of the planetary embryo.}
        \label{fig:models}
\end{figure*}

\subsection{Set-up}

We largely follow the setup of \citet{ChM2021}. Only a brief summary is given here. We consider a 3D patch of
a non-self-gravitating disk. The simulations are performed in a frame co-rotating with the planetary embryo, 
using a polar spherical grid centered on the star.
 The numerical domain extends
for $0.9r_{p}$ to $1.1r_{p}$ in the radial direction, from $-\pi/18$ to $\pi/18$ in
the azimuthal direction and between $\pi/2-h_{p}$ and $\pi/2+h_{p}$ in colatitude (here
$h_{p}=0.05$ is the aspect ratio at the position of the planetary embryo). 
The simulations employ $(N_r,N_\theta,N_\phi)=(880,128,1536)$ grid zones, corresponding to cell sizes of 
$(2.3\times10^{-4}r_p, 7.8\times10^{-4}r_p,2.3\times10^{-4}r_p)$. 

Initially, the surface density and temperature follow constant radial profiles, i.e.
$\Sigma\propto r^{-\sigma}$ and $T\propto r^{-\zeta}$, with $\sigma=\zeta=0$.
The disk is initially threaded by a toroidal magnetic field given by
\begin{equation}
B_\phi=H\Omega\sqrt{\frac{8\pi\rho}{\beta}},
    \label{eq:Bphi}
\end{equation}
whose strength is parameterized by the plasma-$\beta$ parameter, with $\beta = 50$ and $1000$.
Here, $H$ denotes the pressure scale length.

We consider a planetary embryo of mass $M_p$, with a luminosity of $L\in \{0,L_c\}$, where $L_{c}$ is the critical luminosity
\begin{equation}
L_c=\frac{4\pi GM_p\chi\rho_0}{\gamma},
 \label{eq:luminosity}
\end{equation}
with $\rho_0=\Sigma_0/(\sqrt{2\pi}H)$ the unperturbed
midplane density at $r=r_p$ \citep{Masset2017}. 
For $L>0.94L_{c}$, heat release dominates over thermal diffusion and the total thermal torque is positive, 
whereas for $L<0.94L_{c}$ diffusion dominates and the total thermal torque is negative. At $L=0.94L_{c}$, 
the thermal contribution vanishes (see Fig. 6 \citeauthor{ChM2021} \citeyear{ChM2021}). We set thermal diffusivity to $\chi=10^{-5}r_p^2\Omega_p$. 

Since planetary embryos are the seeds from which larger planets grow, a natural first step is to investigate
whether the thermal torques that govern their migration in laminar disks (where they can slow down or even reverse migration and
thereby favor their growth) remain effective in a turbulent medium. Therefore, unless otherwise stated, we adopt $M_{p}=0.33M_{\mars}$, with $M_{\mars}$ the mass of Mars. In physical units,
further adopting $r_p =5.2\,\mathrm{au}$ and $\Sigma_0=208\,\mathrm{g\,cm^{-2}}$
yields $\chi=1.1\times 10^{20}$ cm$^{2}$ s$^{-1}$ and $L_c=7.8\times10^{25}\mathrm{erg}\,\mathrm{s}^{-1}$. The value of $\chi$ appears reasonable
at this distance from the central star. In principle, the results of our simulations can be rescaled to a different orbital radius $r_{p}$. In that case, the thermal diffusivity scales as
$\chi=1.1\times 10^{20} (r/5.2\,{\rm au})^{1/2}{\rm cm^{2}\,s^{-1}}$. At large radii, however,
this simple scaling is expected to deviate from the true value because 
the radial dependencies of the opacity \citep{BL1994}, density, and temperature are unlikely to conspire to produce such a scaling of the diffusivity. Therefore, caution must be exercised when rescaling the simulation results to other orbital distances.

\subsection{Code and boundary conditions}

To solve the continuity, momentum and energy equations for the gas, we use the hydrodynamic public \textsc{Fargo3D} multifluids code \citep[][]{BLlKP2019}, which is an extended version of the original \textsc{Fargo3D} code \citep[][]{BLlM2016}, with orbital advection enabled \citep[][]{Masset2000}. 

A thermal diffusion module was included in the code {\textsc{Fargo3D}\footnote{The version of this code without the thermal diffusion module is available in https://fargo3d.github.io/documentation/}} multifluids. This module was already tested and used in \citet{ChM2021} for the numerical study of thermal torques in a gaseous disk without magnetic fields. Using an operator-splitting scheme, the thermal diffusion term in the energy equation was evolved separately. Specifically, during the diffusion update the internal energy satisfies
\begin{equation}
\partial_te=-\nabla\cdot\mathbf{F}_\mathrm{H}.
    \label{eq:energyflux}
\end{equation}
The source energy term $S_{p}$, representing
the heat release of the planetary embryo, was implemented by increasing the internal energy density 
over eight adjacent grid cells
\citep[]{BLl_etal2015,ChM2021,VR2024} 
\begin{equation}
\Delta e_{i,j,k}=L\left(1-\frac{\phi_{i,j,k}^p}{\Delta \phi}\right)\left(1-\frac{r_{i,j,k}^p}{\Delta r}\right)\left(1-\frac{\theta_{i,j,k}^p}{\Delta \theta}\right)\frac{\Delta t}{V_{i,j,k}},
    \label{eq:eight_cm}
\end{equation}
where $L$ is the luminosity, $V_{i,j,k}$ the volume of the cell $(i,j,k)$, $\phi_{i,j,k}^p=|\phi_p-\phi_{i,j,k}|$, $r_{i,j,k}^p=|r_p-r_{i,j,k}|$, $\theta_{i,j,k}^p=|\theta_p-\theta_{i,j,k}|$ and $\Delta t$ is the difference between each timestep. 

The boundary conditions follow those implemented for the gas density and velocity components in \citet[][]{ChM2021}. We include magnetic resistivity buffer zones in the radial and vertical directions of the disk, covering the same computational domain as the hydrodynamical damping zones. 

\subsection{Parameters and relevant scales within thermal torques theory}

Linear theory predicts that the size of the thermal disturbance is given by 
\begin{equation}
\lambda_c=\sqrt{\frac{\chi}{(3/2)\Omega_p\gamma}}
 \label{eq:lambda}
\end{equation}
\citep{Masset2017}.
The level of asymmetry of the thermal perturbations is determined by the planetary embryo's corotation offset, which is given by
\begin{equation}
x_p=\eta h_p^2r_p,
 \label{eq:corotation}
\end{equation}
where $\eta$ is a function of the power-law exponents of the surface density and temperature, given by
\begin{equation}
\eta=\frac{\sigma}{3}+\frac{\zeta+3}{6}.
\end{equation}
In our case, for $\sigma=\zeta=0$, the parameter $\eta$ reduces to $1/2$.

\begin{figure*}
\includegraphics[width=\textwidth]{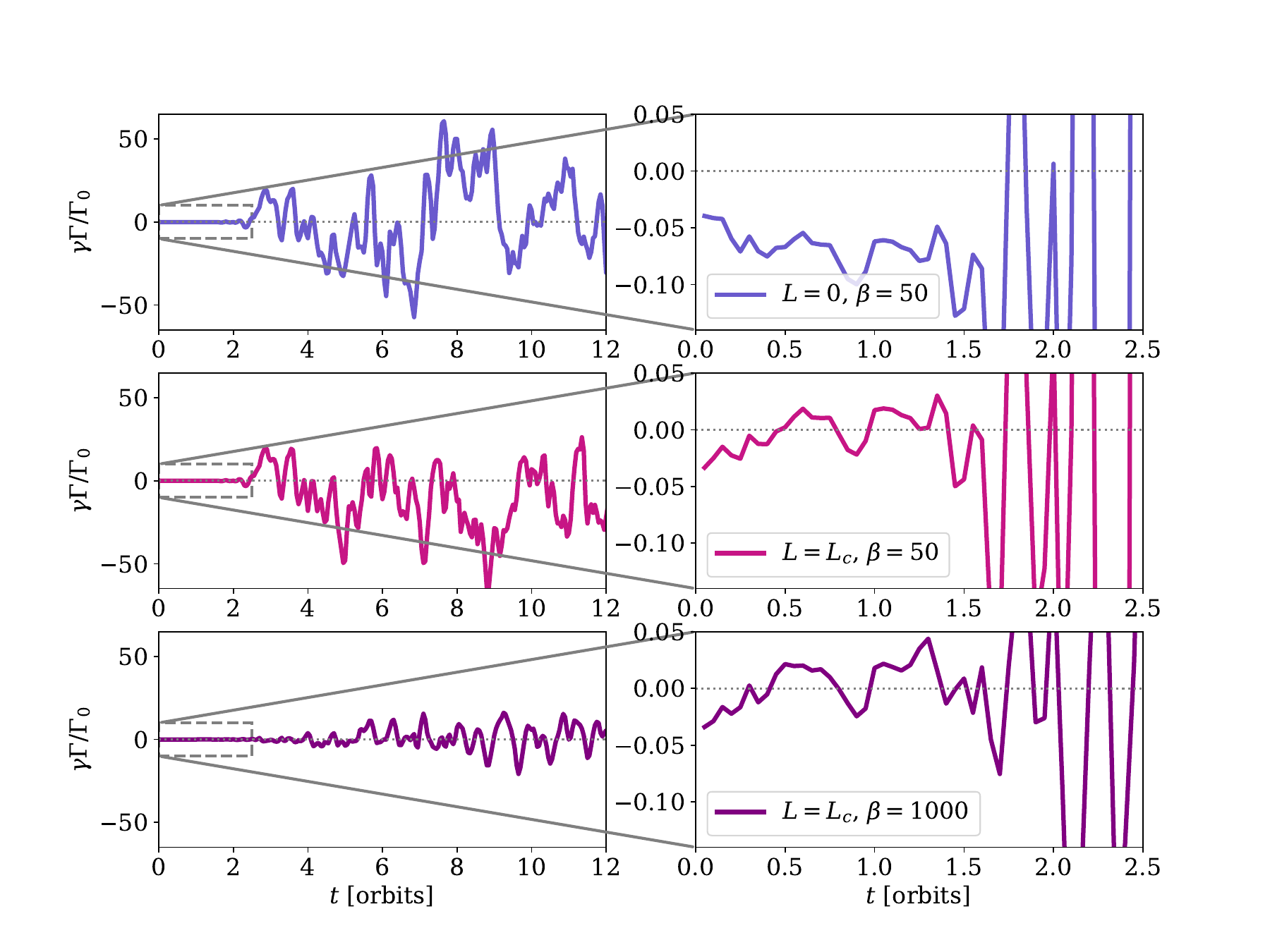}
 \caption{Temporal evolution of the total torque $\Gamma$ (in units of $\Gamma_0/\gamma$) on the planetary embryo.}
\label{fig:torq}
\end{figure*}

\section{Results}
\label{sec:results}

\subsection{The turbulent disk}
For the adopted disk parameters, Eqs. (\ref{eq:lambda}) and (\ref{eq:corotation}) yield $x_{p}=1.25\times 10^{-3}r_{p}$, $\lambda_{c}=2.2\times 10^{-3}r_{p}$,
and $H=0.05r_{p}$. The ordering $x_p\leq\lambda_c\ll H$ implies that 
the density perturbation induced by the planetary embryo's luminosity  
should be observable, at least in the absence of turbulence or 
prior to its full development, which occurs after approximately three orbital periods. To verify
this expectation, 
Fig. \ref{fig:lobes} presents the density perturbation for $L=L_c$ and $\beta=50$ after $t=1$ orbit, i.e. before turbulence has had time to grow. At this stage, the hot lobes are already
fully formed and exhibit a pronounced asymmetry with respect to the planetary embryo's position. 
This behavior is characteristic of the evolution of hot lobes in a turbulence-free gas disk \citep[][]{ChM2021}.

The onset of turbulence is illustrated in
Fig. \ref{fig:alpha}, which shows the temporal evolution of the stress-to-pressure ratio $\alpha$, a measure of 
the strength of the turbulence in the disk, computed as

\begin{equation}
\alpha(r,t)=\dfrac{\int (\rho \,\delta v_r\,\delta v_\phi-B_r B_\phi/4\pi)\,r^2\sin\theta \,d\theta \,d\phi}{\int p \,r^2\sin\theta \,d\theta\, d\phi},
  \label{eq:alpha}
\end{equation}
for models with $L=L_c$ and for the two values $\beta=50$ and $\beta=1000$. 
The development of MRI-driven turbulence begins at $t\sim1.8$ orbits, and by $t=3$ orbits $\alpha$ already exceeds $10^{-4}$. Note that when turbulence is fully developed, $\alpha$ reaches the values of $10^{-3}$ and $10^{-2}$, when $\beta=1000$ and $\beta=50$, respectively. 

To ensure that the MRI is well resolved in our models, we compute the quality factor $Q_{\phi}$ \citep[][]{Sor2012,Flock2017}
\begin{equation}
Q_\phi=2\pi\sqrt{\frac{16}{15}}\frac{|B_\phi|}{\sqrt{4\pi\rho}}\frac{1}{\Omega}\frac{1}{r\Delta \phi},
    \label{eq:Qf}
\end{equation}
for the model with a weak magnetic field ($\beta=1000$). Fig. \ref{fig:qf} shows the radial profile of $Q_\phi$ (temporally and spatially $\{\theta,\phi\}$ averaged). 
The fact that $Q_{\phi}>8$ indicates that the MRI is well resolved. 

Because turbulence causes fluid elements in the vicinity of the planetary embryo (and more generally in the disk) to follow different trajectories in different realizations, a direct subtraction between simulations with $L=L_{c}$
and $L=0$ does not provide useful information. We therefore consider the density maps without subtracting the cold case. As shown in Fig. \ref{fig:models}, independently of the magnetic field strength, turbulent fluctuations generate cells of high and low density (i.e. strong density gradients) in the same area that would otherwise be occupied by hot lobes (see Fig. \ref{fig:lobes}). Turbulence thus effectively wipes out the thermal lobes.

\subsection{Turbulence suppresses heating torques}
\label{sec:discussion}
The activation time of thermal torques is about one orbital period of the planetary embryo \citep[][]{ChM2021}, whereas,
as discussed above, the onset of MRI-driven turbulence occurs after about $2$ orbital periods.
In line with this, Fig. \ref{fig:torq} shows that, for times $t\leq1.5$ orbits, the total torque $\Gamma$ follows the behavior expected for 
the thermal torque in a laminar gas disk. During this phase, the torque is negative for $L=0$, as expected for a 
non-luminous planetary embryo, whereas it becomes positive when $L=L_c$\footnote{Luminosities above the critical luminosity produce  
smaller changes in the total torque (i.e. they display an offset nearly constant
in time between them, see Fig. 5 in \citet{ChM2021}) than the strong oscillations reported here when the MRI is active.}. Importantly, the torque behavior is largely insensitive
to the value of $\beta$ over this time interval (see the zoomed region in the right column of Fig. \ref{fig:torq}),
reflecting the fact that turbulence has not yet developed in the disk.

For $t>1.5$ orbits, the situation changes drastically: the torque increases substantially in all cases and exhibits a strongly oscillatory behavior, reflecting the onset of the turbulence in the disk. A careful comparison of the torque in the cases where $\beta=50$ and $L=0$ and $L=L_c$ (see first and second rows in Fig. \ref{fig:torq}) suggests that the torque falls into the stochastic regime \citep{NP2004,Nelson2005,BL2010,Comins_etal2016}. In the case $L=L_{c}$, the torque is shifted to higher (more positive) values, while the overall shape of $\Gamma(t)$
remains largely unchanged, even up to $\sim 3$ orbits. Furthermore, when the value of $L$ is held fixed and $\beta$ is increased from $50$ to $1000$, the amplitude of the torque oscillations is reduced (see the bottom row of Fig. \ref{fig:torq}). This indicates that the strength of the magnetic field governs the stochastic nature of the torque acting on the planetary embryo.

We expect the stochastic nature of the torque to be largely insensitive 
 if the luminosity or mass of the planetary embryo are increased, since thermal torques become significantly cut off once the inequality $(\lambda_c/H)M_\mathrm{th}< M_p< M_\mathrm{th}$ is satisfied, where $M_\mathrm{th}=c_s^3/G\Omega_{p}$ is the thermal mass \citep{VRM2020,ChM2021}.

\subsection{No traces of active magnetic resonances}
 
When the toroidal magnetic field is sufficiently strong ($\beta \lesssim 4$), the flow remains in a laminar regime, and magnetic resonances can make a significant contribution to the total torque.
These resonances occur
where the Doppler-shifted perturbation frequency in the fluid frame matches that of a slow MHD wave propagating along the magnetic field lines \citep{Terquem2003}. 
The relative distance of these magnetic resonances from the planetary embryo is determined by the strength of the toroidal magnetic field through the $\beta$-parameter \citep{Terquem2003,Uribe2015}:
\begin{equation}
|r_\mathrm{M}-r_p|=\frac{2H_{p}}{3\sqrt{1+\beta_{p}}},
    \label{eq:mr}
\end{equation}
where the subscript $p$ indicates that the corresponding quantity should be evaluated at the position of the planetary embryo. As expected, when the magnetic field is very weak $\beta_{p}\to\infty$, the position of the magnetic resonances converges to the position of the planetary embryo. These resonances can dominate the Lindblad torques, as they are located closer to the embryo than the Lindblad resonances.

To assess whether magnetic resonances operate within the thermal lobes prior to the
onset of turbulence, we compare the radial separation
$|r_M - r_p|$ with the characteristic cutoff scale $\lambda_c$. In our models,
$|r_\mathrm{M}-r_p|=4.7\times10^{-3}r_{p}$ and $|r_\mathrm{M}-r_p|=1.05\times10^{-3}r_{p}$, for $\beta=50$ and $\beta=1000$, respectively. On the other hand, 
$\lambda_c=2.2\times10^{-3}$ (from Eq. \ref{eq:lambda}). Thus, $\lambda_c$ and $|r_\mathrm{M}-r_p|$ are
comparable.
In our models, however, we find no evidence that magnetic resonances are active in the vicinity of the planetary embryo. Note that the timescale for the development of large-scale density waves associated with magnetic resonances is of the order of a few orbital periods \citep{Uribe2015}. Consequently, additional perturbations beyond those produced by the thermal lobes may not appear within the single orbital period shown in Fig. \ref{fig:lobes}. 

In the case of a relatively weak magnetic field ($\beta\gg 1$), as in the models considered in this work,
we have seen that the disk becomes turbulent.  In principle, turbulence can suppress the heating torques, whereas the magnetic torques may persist at least partially.
Returning to Fig. \ref{fig:models},
we find no evidence of MHD ring-like waves such as those reported by \citet[][see their figure 3]{Frog2005}. We emphasize that the resolution used in our simulations is sufficient to resolve the
characteristic density perturbation $\lambda_c$ induced by planetary heating with at least ten grid cells, and therefore is also adequate to resolve the radial separation $|r_{\mathrm M} - r_p|$. 
This indicates that the absence of slow MHD waves is not a resolution effect, but rather a consequence of the turbulent state of the disk, in which the magnetic field fluctuates locally on timescales shorter than those required for slow MHD waves to develop \citep[see also][]{Terquem2003,Comins_etal2016}. 
This result complements what was previously reported for an isothermal disk with a toroidal magnetic field,
in which magnetic resonances were likewise found to be ineffective \citep[][]{Uribe2015}.

\subsection{Thermal torques for Earth-mass planetary cores}
\begin{figure}
    \centering
    \includegraphics[width=1.0\linewidth]{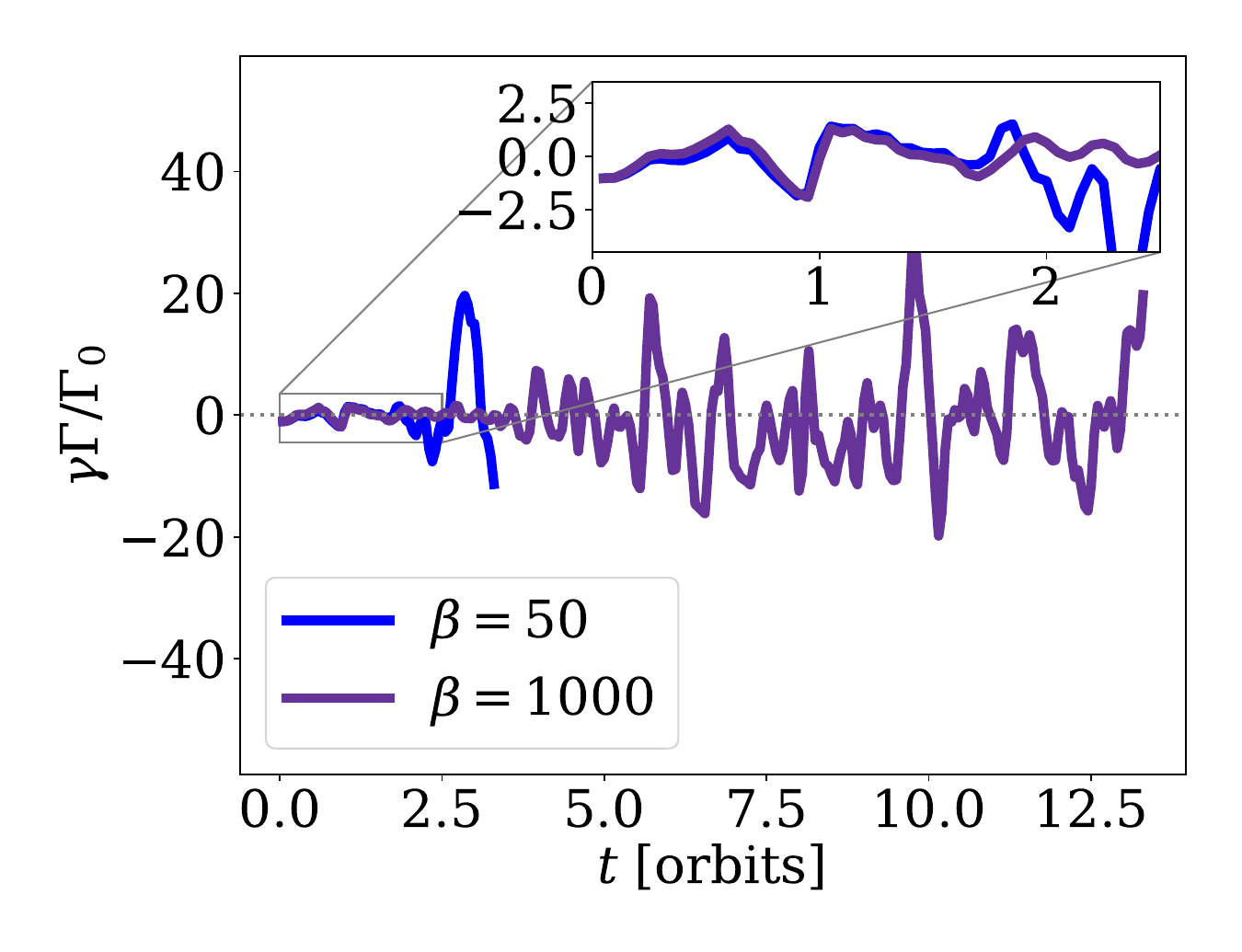}
    \caption{Temporal evolution of the total torque $\Gamma$ (in units of $\Gamma_0/\gamma$) on a planet of mass $M_p=1\,M_\oplus$, for $\beta=50$ (blue lines) and $\beta=1000$ (purple lines).}
    \label{fig:tq1ME}
\end{figure}

In the previous subsections, we focused on the survival of thermal
torques for planetary embryos with masses close to that of Mars in
a turbulent disk. Since thermal torques reach their maximum efficiency
for planetary cores of order one Earth mass \citep{VRM2020}, 
we ran two additional models with $M_p=1\,M_\oplus$ to assess whether
thermal torques can survive in this mass range.
Fig. \ref{fig:tq1ME} shows the resulting torques for $L=L_{c}$ and two values of $\beta$
($50$ and $1000$)\footnote{The $\beta=50$ model evolved somewhat more slowly;
due to the high computational cost, it was only integrated up to $t=3.3$ orbits, the limit imposed by cluster's walltime. Nevertheless, this duration was sufficient to capture the abrupt change in torque resulting from the onset of MRI.}. 
As in the case of the planetary embryo, after $1.5$ orbits turbulence induces strongly oscillatory torques for both values of $\beta$. We do not show maps of the gas density
as they are very similar to those presented in Figure \ref{fig:models}.

Large torque fluctuations, very similar to those shown in Figure \ref{fig:tq1ME},
have also been observed in simulations of super-Earth planets $(M_p>3\,M_\oplus)$.
In that case, they arise from vortices forming at the edge of the horseshoe region \citep[][]{ChM2021},
as the gas dynamics become highly nonlinear with the Hill sphere for such planetary masses.
Therefore, in regions where MRI turbulence develops, migration is expected to be stochastic for
$0.03M_{\oplus}\lesssim M_{p}\lesssim 1M_{\oplus}$ as a result of MRI turbulence and
for $M_{p}\gtrsim 3M_{\oplus}$ owing to vortices.

Finally, we stress that the turn-off effect on the torque reported here  is 
distinct from the cut-off effect investigated in \citet{VRM2020}.

\section{Towards more complex models}
\label{sec:complex_mod}

In our simulations, we adopted a constant thermal diffusivity $\chi=10^{-5}r_p^2\Omega_p$. This
value was chosen to allow the formation of hot
thermal lobes around the luminous planet, enabling us to quantify the heating torques and construct a thermally controlled environment prior to the onset of MRI. In real protoplanetary disks, however, $\chi$ is expected to vary spatially as it depends on local physical conditions, notably the gas density, temperature, and the opacity of the
medium \citep[][and references therein]{JM2017}.  
These quantities are, in turn, influenced by the local dust concentration and by dust
properties such as the grain size distribution and composition \citep{Chame2025}. Therefore, more sophisticated models that include non-constant thermal diffusivity and an explicit treatment of dust are warranty and will be explored in future work.

\section{Conclusions}
\label{sec:conclusions}

In this study, we report the results of  high-resolution 3D MHD simulations of a planetary embryo with a mass one third that of the planet Mars and for an Earth-mass core, embedded in a stratified gaseous disk with thermal diffusivity and a toroidal magnetic field. Our aim was to assess whether thermal torques can drive planetary migration in turbulent regions of a protoplanetary disk, such as the inner and outer regions of the disk located outside the so-called dead zone. To this end, we have considered two different luminosities for 
both the planetary embryo and core: $L=0$ corresponding to a cold thermal torque, and $L=L_c$, corresponding to a heating torque. In all cases, we find that, as turbulence develops in the disk,  
the thermal lobes are disrupted for both $\beta=50$ and $\beta=1000$. In this turbulent regime, the total torque acting on the Mars planetary embryo and Earth-sized core displays the characteristic pattern of a stochastic torque, implying that its migration should proceed in the stochastic migration regime, similar to that observed in \citet{NP2004} and \citet{Comins_etal2016}. Given the results of \citet{ChM2021},
which show that larger planetary masses also experience stochastic torques, we conclude that planets with masses $0.03M_{\oplus}\lesssim M_{p}\lesssim 1M_{\oplus}$ and $M_{p}\gtrsim 3M_{\oplus}$ are insensitive to thermal torques in a disk where turbulence is driven by MRI or vortices. Lastly, in agreement with the results of \citet{Comins_etal2016}, we found that the level of turbulence
attained for $\beta\geq 50$ renders magnetic resonances ineffective.

\begin{acknowledgements}
We thank the referee for the careful reading of the manuscript and for the constructive comments that helped improve the paper.
     The work of R.O.C. was supported by the Czech Science Foundation (grant 25-16507S). Computational resources were available thanks to the Ministry of Education, Youth and Sports of the Czech Republic through the e-INFRA CZ (ID:90254). Also, R.O.C. thanks Neal J. Turner for the fruitful discussions on this manuscript.
\end{acknowledgements}

\bibliographystyle{aa} 
\bibliography{example} %

\end{document}